\documentclass[11pt,fleqn]{article}       
\usepackage{geometry}
\usepackage[T1]{fontenc}
\usepackage[utf8]{inputenc}
\usepackage{authblk}
\usepackage{mathrsfs}
\usepackage[font=small,labelfont=bf]{caption}
\usepackage{graphicx}
\usepackage{multirow}
\usepackage{amsmath,amssymb,amsfonts}
\usepackage{verbatim}
\usepackage{bm}
\usepackage{color}
\usepackage{ulem}
\usepackage{mathtools}
\usepackage{cite}
\usepackage{colortbl}
\definecolor{light-gray}{gray}{0.85}
\usepackage[pdftex,colorlinks,pdfpagelabels]{hyperref}
\hypersetup{
   bookmarksnumbered,
   citecolor={blue},
   linkcolor={blue},
   urlcolor={blue},
   filecolor={blue}
} 
\newcommand{\eVdist}{\kern-0.06em}

\newcommand{\gev}{\:\text{Ge\eVdist V}}

\newcommand{\D}{\mathrm{d}}
\newcommand{\position}{r}

\newcommand{\AddrTexas}{%
\textit{Department of Physics, The University of Texas at Austin, Austin, 78712 TX, USA}
}
\newcommand{\AddrStockholm}{
\textit{Oskar Klein Center for Cosmoparticle Physics, University of Stockholm, 10691 Stockholm, Sweden}
}
\newcommand{\AddrNordita}{
\textit{Nordita, KTH Royal Institute of Technology and Stockholm University, Roslagstullsbacken 23, 10691 Stockholm, Sweden}
}

\usepackage{fancyhdr}
 
\fancypagestyle{plain}{%
    \fancyhead[R]{NORDITA-2020-107\\UTTG 22-2020}
    
}
\date{}
\title{\Large\bf The Power Spectrum of Density Perturbations in Chain Inflation}
\author[1,2]{Martin Wolfgang Winkler\thanks{martin.winkler@su.se}}
\author[1,2,3]{Katherine Freese\thanks{ktfreese@utexas.edu}}
\affil[1]{\AddrTexas}
\affil[2]{\AddrStockholm}
\affil[3]{\AddrNordita}

\begin{document}
\maketitle
\vspace*{0mm}
\begin{abstract}
Chain Inflation is an alternative to slow roll inflation in which the universe undergoes a series of transitions between different vacua. The density perturbations (studied in this paper) are seeded by the probabilistic nature of tunneling rather than quantum fluctuations of the inflaton. We find the scalar power spectrum of chain inflation and show that it is fully consistent with a $\Lambda$CDM cosmology. 
In agreement with some of the previous literature (and disagreement with others), we show that $10^4$ phase transitions per e-fold are required in order to agree with the amplitude of Cosmic Microwave Background anisotropies within the observed range of scales.
Interestingly, the amplitude of perturbations constrains chain inflation to a regime of highly unstable de Sitter spaces, which may be favorable from a quantum gravity perspective since the Swampland Conjecture on Trans-Planckian Censorship is automatically satisfied. We provide 
new analytic estimates 
for the bounce action and the tunneling rate in periodic potentials which replace the thin-wall approximation in the regime of fast tunneling. Finally, we study model implications and derive an upper limit of $\sim 10^{10}$ GeV on the axion decay constant in viable chain inflation with axions.
\end{abstract}
\clearpage

\section{Introduction}

Cosmic inflation~\cite{Guth:1980zm} is one of the corner stones of modern cosmology. The rapid expansion of space solves the horizon, flatness and monopole problems. At the same time, it seeds density perturbations in the primordial plasma which we observe as the temperature fluctuations of the Cosmic Microwave Background (CMB). 

In Guth's original proposal inflation occurs due to a false vacuum state in which the universe is initially trapped. The energy density of the false vacuum drives the quasi-exponential expansion of space. During inflation, tunneling processes lead to the formation of bubbles within the sea of false vacuum. The insides of these bubbles reside in the true vacuum state, while the energy released by the transition is stored in the bubble walls. Proper transition of the universe from inflation to the radiation dominated phase requires that the bubbles percolate and release their energy through particle production. This is where Guth's `old inflation' model fails: in order for inflation to solve the mentioned problems of big bang cosmology it has to last for at least $\sim 60$ e-folds. As a consequence, the tunneling rate between the vacuum states needs to be suppressed. Bubbles would only form very distantly from each other and never collide - even with bubble walls expanding at the speed of light. The insides of the bubbles would forever stay cold and empty -- a universe which looks very different from what we observe~\cite{Guth:1982pn}.

The problem of Guth's `old inflation' was soon resolved by replacing the picture of vacuum tunneling by that of a scalar field slowly rolling down its potential~\cite{Linde:1981mu,Albrecht:1982wi}. In slow roll inflation, reheating is no longer connected to bubble collisions. It, rather, occurs when the inflaton decays into a hot bath of particles. 

However, there also exists a completely different approach to rectify Guth's original model\footnote{Another approach 
to a successful tunneling model of inflation is Double Field Inflation~\cite{Adams:1990ds,Linde:1990gz}.  In this model, the potential is multidimensional: one direction requires tunneling to get from the false to the true vacuum; in another direction the field rolls. The
Universe is initially stuck in a false vacuum for a long time.  Yet, as the field moves in the rolling direction, the barrier to tunneling becomes much smaller, so that at some point the tunneling rate switches from very slow to very fast, and 
the Universe reheats suddenly and uniformly.}
: in `chain inflation'~\cite{Freese:2004vs,Freese:2005kt} the universe contains a series of false vacuum states of different energy instead of just one. The existence of multiple metastable vacua was originally motivated by the string landscape~\cite{Freese:2004vs,Freese:2006fk}. But theories with axions also naturally give rise to the desired vacuum structure if they contain softly broken discrete shift symmetries~\cite{Freese:2005kt,Ashoorioon:2008pj}. In chain inflation, the universe tunnels through a series of vacuum states of ever lower energy. If there occur at least three tunnelings per e-folding, the corresponding vacuum bubbles are generated sufficiently close to percolate and thermalize~\cite{Guth:1982pn,Turner:1992tz} so that reheating is successful.

While it is encouraging that chain inflation avoids the failure of old inflation, additional constraints must be satisfied by a successful theory of the early universe. Specifically, we need to verify that chain inflation produces the observed pattern of CMB temperature fluctuations. For this purpose, the scalar power spectrum of chain inflation has to be determined. One might naively think that scalar perturbations, as in the case of slow roll inflation, are linked to the quantum fluctuations of the inflaton. Instead, in chain inflation, they  originate from (i) the probabilistic nature of tunneling which places separate locations in different vacuum states and (ii) the bubble collisions during the percolation and thermalization process. In this paper we study the first of these two mechanisms, namely the perturbations arising from the probabilistic nature of tunneling.

While several previous calculations of scalar fluctuations in chain inflation based on this mechanism exist~\cite{Watson:2006px,Feldstein:2006hm,Huang:2007ek,Chialva:2008zw,Chialva:2008xh,Cline:2011fi}, these are in mutual disagreement. We, therefore, decided to clarify the situation and rederive the scalar power spectrum. We will show that a nearly scale invariant spectrum, consistent with $\Lambda$CDM cosmology, is realized if tunneling and Hubble rate are slowly varying functions of time. The amplitude of CMB anisotropies will allow us to determine the tunneling rate in chain inflation. We will then turn to model realizations.

For the case of a periodic potential (tilted cosine), we present new analytic approximations of the bounce action and tunneling rate which were obtained by a fit to our numerical results (using Coleman's formalism~\cite{Coleman:1977py}). 
These expressions constitute a significant improvement compared to the widely used thin-wall approximation (which badly fails in the regime of fast tunneling).
Our calculation of the tunneling rate will allow us to directly apply the CMB constraints to axion models of chain inflation and to obtain a constraint on the axion decay constant.

\section{Basics of Chain Inflation}

\begin{figure}[h!]
\begin{center}
  \includegraphics[width=\textwidth]{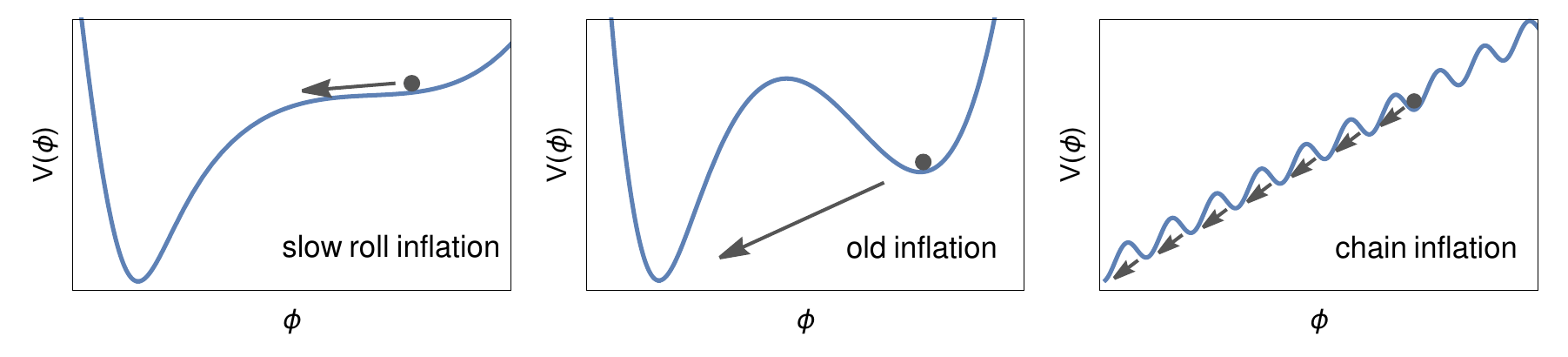}
\end{center}
\vspace{-6mm}
\caption{Illustration of inflation theories. In slow roll inflation, a scalar field slowly moves down its potential. In old inflation, the universe undergoes a single phase transition. In chain inflation, the universe tunnels along a chain of vacua with slowly decreasing energy.} 
\label{fig:inflationmodels}
\end{figure}

Chain inflation and slow roll inflation are two fundamentally different theories of the early universe. While both resolve the problems of old inflation, Guth's proposal of the universe trapped in a false vacuum state is dismissed within the slow roll paradigm. Chain inflation, on the other hand remedies the tunneling picture: the universe undergoes a series of transitions towards lower and lower energy vacua (see Fig.~\ref{fig:inflationmodels} for an illustration of the the different inflation theories). 

As discussed in the introduction, reheating in Guth's old inflation fails because of the slow tunneling rate required
to achieve 60 e-folds of inflation; the bubbles that are nucleated by the phase transition remain far apart and empty rather than converting to the radiation domination required for our Universe to proceed.
Chain inflation, on the other hand, is a series of rapid tunneling events from one minimum to another.
In each step of the chain,
the phase transition is sufficiently rapid that bubbles of the new vacuum do indeed percolate.
As the bubbles collide, the energy of the bubble walls is converted to radiation and the phase 
transition completes.  Since the field is trapped in each minimum for only a short time, the amount of inflation per
step is small, and many steps are required.
 
Here we will review how fast the tunneling must be for percolation to succeed, as shown by Guth and Weinberg~\cite{Guth:1982pn}. An approximate bound on the bubble nucleation rate can be derived by taking the Hubble parameter $H$ to be constant such that the scale factor of the universe becomes $a(t)\propto e^{H t}$. 

Since bubble walls expand approximately at the speed of light, the volume of a single bubble created at time $t_0$ can be approximated as~\cite{Guth:1982pn}
\begin{equation}
 V_{b}\simeq \frac{4\pi}{3 H^3} e^{3H(t-t_0)}\,,
\end{equation}
where $t\gg t_0$ was assumed. 
Bubbles are continuously nucleated at the rate per unit four-volume $\Gamma$. If $\Gamma$ is approximated as a constant, the volume filled by all bubbles compared to the total volume is~\cite{Guth:1982pn}
\begin{equation}
    \frac{\sum V_{b}}{V_{\text{tot}}} \simeq  \frac{4\pi}{9}\frac{\Gamma}{H^4}\,.
\end{equation}
If this number is larger than one, bubbles necessarily overlap and successful percolation can be achieved. Colliding bubble walls can transfer their energy into radiation and the universe is reheated properly. The following constraint on the bubble nucleation rate per Hubble four-volume is obtained
\begin{equation}
  \frac{\Gamma}{H^4} \gtrsim \frac{9}{4\pi}\,.
\end{equation}
Bubble percolation, hence, requires >$\mathcal{O}(1)$ vacuum transitions per e-fold of inflation (i.e., less than an e-fold
of inflation per tunneling event). In order to solve the problems of Big Bang cosmology, inflation requires at least 60 e-folds of inflation.  Thus chain inflation must feature at least $\mathcal{O}(100)$ vacuum transitions. We will show in this work that an even larger number is required in order to generate the observed magnitude of CMB temperature fluctuations.

\section{Density Perturbations}
In order to relate chain inflation to physical observables in the CMB, we need to know the power spectrum of the primordial scalar perturbations.
\subsection{Previous Results}
 At this point, we wish to stress the results from previous derivations of the scalar power spectrum $\Delta_{\mathcal{R}}^2$ of chain inflation
\begin{align}\label{eq:powerprevious}
\Delta_{\mathcal{R}}^2 &= \frac{H^2}{4\pi^2 c_s \epsilon}(-k c_s \eta)^{-2\epsilon} & \text{(Watson et al.~\cite{Watson:2006px})\,,}\nonumber\\
\Delta_{\mathcal{R}}^2 &= (0.04\pm 0.02) \left(\frac{\Gamma}{H^4} \right)^{-0.42\pm 0.03} & \text{(Feldstein \& Tweedie~\cite{Feldstein:2006hm})\,,}\nonumber\\
 \Delta_{\mathcal{R}}^2 &= \frac{H^2}{8\pi^2\epsilon} & \text{(Huang~\cite{Huang:2007ek})\,,}\nonumber\\
 \Delta_{\mathcal{R}}^2 &= \frac{H^2}{8\pi^2\epsilon/\sqrt{3}}\;\;\;´& \text{(Chialva \& Danielsson~\cite{Chialva:2008zw,Chialva:2008xh})\,,}\nonumber\\
 \Delta_{\mathcal{R}}^2 &= \frac{3}{4\pi} \frac{H^4}{\Gamma} & \text{(Cline et al.~\cite{Cline:2011fi})}\,,
\end{align}
where $\epsilon=-\dot{H}/H^2$ and $c_s$, $\eta$ denote sound speed, conformal time. It can be seen that the calculations are in vast disagreement. Therefore, we decided to rederive the scalar power spectrum and to clarify the situation.

\subsection{Two-Point Correlation}
In slow roll inflation, density perturbations originate from quantum fluctuations in the inflation field. Such fluctuations are suppressed in chain inflation due to the inflaton mass in each vacuum along the chain. However, there exists a complementary origin of fluctuations related to the probabilistic nature of tunneling.\footnote{While in this work we focus on density perturbations from tunneling, we note that additional perturbations can be created by the collision of bubble walls which separate different vacuum states.} The number of vacuum transitions varies among different spatial locations. The corresponding scalar power spectrum can be derived from the two-point correlation in the inflaton field value.

For the moment, we assume that the tunneling rate per four-volume $\Gamma$ as well as the Hubble rate during chain inflation are both constant. These assumptions will lead to a scale-invariant scalar power spectrum. We will later comment on deviations from scale invariance due to (slow) variations in $\Gamma$ and $H$.

We take inflation to start at the time $t=0$ from an initially homogeneous patch with constant field value $\phi(\position,0)=\phi_0$. (This choice of initial condition is not crucial since the universe can otherwise be homogenized by a few e-folds of inflation.) As usual, we introduce the comoving position variable $\position$ which is obtained from the position in real space by dividing out the scale factor of the universe $a(t)$. For convenience, we define $a(0) = 1$ such that
\begin{equation}\label{eq:scalefactor}
a(t) = e^{H t}\,,
\end{equation}
for our assumptions of constant $H$ and $\Gamma$.

Now, we turn to the two-point correlator $\langle \delta\phi(\position,t)\delta\phi(0,t)\rangle$ of inflaton field fluctuations, where we define 
\begin{equation}
\delta\phi(\position,t) =\phi(\position,t)-\langle\phi(t)\rangle\,.
\end{equation}
Note that the mean inflaton field value $\langle\phi(t)\rangle$ does not depend on the spatial location. We are interested in the correlator, when the comoving distance scale crosses the horizon, $|\position|= (aH)^{-1}$. The time of horizon crossing shall be denoted by $t_*$ in the following. Since the correlator is preserved on super-horizon scales (causally disconnected patches evolve independently) we can equivalently look at the correlator at any later time $t>t_\star$ within the inflationary epoch.

\begin{figure}[h!]
\begin{center}
  \includegraphics[width=7cm]{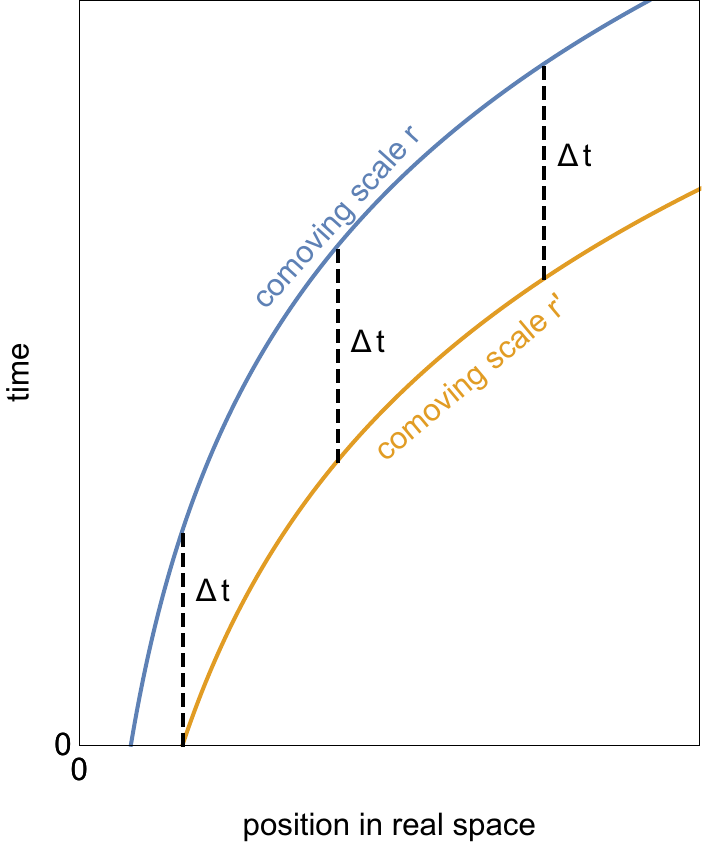}
\end{center}
\vspace{-6mm}
\caption{During inflation the two comoving scales $\position$ and $\position^\prime$ evolve in the same way, only shifted by the time $\Delta t$.} 
\label{fig:comoving}
\end{figure}

Notice that the correlators of two different comoving scales $\position$ and $\position^\prime$ are related in a simple way. Due to the expansion of the universe the physical size $a(t)\position$ equals the physical size $a(t-\Delta t)\position^\prime$ at a different time $t-\Delta t$ (see Fig.~\ref{fig:comoving}). Hence, the correlators corresponding to $\position$ and $\position^\prime$ evolve in the same way, only shifted by the time $\Delta t$. This is a manifestation of scale-invariance. Taking $|\position|<|\position^\prime|$ for concreteness, we obtain the relation
\begin{align}\label{eq:scale-invariance}
  \langle \delta\phi(\position,t)\delta\phi(0,t) \rangle 
     =&\;\langle \delta\phi(\position,\Delta t)\delta\phi(0,\Delta t) \rangle 
     + \langle \delta\phi(\position,t)\delta\phi(0,t) \rangle  - \langle \delta\phi(\position,\Delta t)\delta\phi(0,\Delta t) \rangle   \nonumber\\
    =&\;\langle \delta\phi(\position,\Delta t)\delta\phi(0,\Delta t) \rangle 
     + \langle \delta\phi(\position^\prime,t-\Delta t)\delta\phi(0,t-\Delta t) \rangle  \nonumber\\
  \xrightarrow{t\gg t_*}&\;\langle \delta\phi(\position,\Delta t)\delta\phi(0,\Delta t) \rangle 
     + \langle \delta\phi(\position^\prime,t)\delta\phi(0,t) \rangle\,.
\end{align}
In the second step we shifted the last two terms by $\Delta t$ which does not affect the correlator difference in a scale-invariant universe and used the fact that  $\langle \delta\phi(\position,0)\delta\phi(0,0) \rangle$ vanishes by definition. The time shift involves a change of the comoving coordinate from $\position$ to $\position^\prime$ to compensate the change of scale factor as described above. The last step amounts to taking the super-horizon limit. Once we know the correlator for a specific $\position$, the correlator for all other scales immediately follows from the above relation.

Given that inflation already lasted a few e-foldings when the CMB scales crossed the Hubble horizon, we can safely assume that they were initially contained in the same vacuum bubble. As has been argued in~\cite{Feldstein:2006hm}, the comoving position $\position$ remains in the same vacuum state as the origin $0$ as long as the separation in real space $a|\position|$ is smaller than the typical space-time distance between two tunneling events $\ell = \Gamma^{-1/4}$. The transition occurs at the time $t_\ell$ defined by
\begin{equation}\label{eq:tell}
t_\ell = -\frac{1}{H}\log\frac{|\position|}{\ell}\,.
\end{equation}
For $t\leq t_\ell$ we can set $\delta\phi(\position,t)=\delta\phi(0,t)$. Now, we use~\eqref{eq:scale-invariance} with $\position^\prime = \ell$ and $\Delta t=t_\ell$ to obtain the superhorizon correlator
\begin{align}\label{eq:correlator3}
  \langle \delta\phi(\position)\delta\phi(0) \rangle \equiv\langle \delta\phi(\position,t)\delta\phi(0,t) \rangle \Big|_{t\gg t_*}
  &= \langle\delta\phi(\position,t_\ell)\delta\phi(0,t_\ell) \rangle  +
  \langle \delta\phi(\ell,t)\delta\phi(0,t) \rangle\Big|_{t\gg t_*}
  \nonumber\\
  &= 
  \text{var}\!\left(\phi(t_\ell)\right) +\text{const} 
 = \frac{\D\text{var}\!\left(\phi\right)}{ \D t} t_\ell+ \text{const} \,
    \nonumber\\
   &= -\log|\position|\frac{\D\text{var}\!\left(\phi\right)}{H \D t}  + \text{const}^\prime\,,
\end{align}
where $\text{var}(\phi(t_\ell))=\langle\phi^2(t_\ell)\rangle - \langle \phi(t_\ell) \rangle^2$ stands for the variance. In the 
second to last step we used that, in a scale-invariant universe, the variance grows linearly in time. The derivative of the variance is, hence, time-independent. In the last step we used the expression Eq.(\ref{eq:tell}) for $t_\ell$. The coordinate-independent constant term (denoted by `$\text{const}$' above) does not affect observables. Furthermore, we absorbed the coordinate-independent part of $t_\ell$ by a redefinition of the (irrelevant) constant term.

\subsection{The Scalar Power Spectrum}
In order to translate the two-point correlation to the power spectrum, we first need to calculate the Fourier-transformed correlator
\begin{equation}
  \langle \delta\phi_{k}\delta\phi_{k^\prime}\rangle = (2\pi)^{3}\delta(k+k^\prime)\frac{2\pi^2}{k^3} \frac{\D\text{var}\!\left(\phi\right)}{H \D t}\,.
\end{equation}
Expressing the field fluctuation in terms of the curvature perturbation $\mathcal{R}=H(\D\langle\phi\rangle/\D t)^{-1}\delta\phi$ yields
\begin{equation}\label{eq:RkRk}
  \langle \mathcal{R}_k \mathcal{R}_{k^\prime}\rangle = \left(\frac{\D \langle\phi\rangle}{H \D t}\right)^{-2}(2\pi)^{3}\delta(k+k^\prime)\frac{2\pi^2}{k^3} \frac{\D \text{var}\!\left(\phi\right)}{H\D t}\,.
\end{equation}
By comparing~\eqref{eq:RkRk} with the definition of the scalar power spectrum
\begin{equation}\label{eq:SPdef}
  \langle \mathcal{R}_k \mathcal{R}_{k^\prime}\rangle=(2\pi)^{3}\delta(k+k^\prime)\frac{2\pi^2}{k^3} \Delta_{\mathcal{R}}^2\,,
\end{equation}
we finally obtain
\begin{equation}\label{eq:powerspectrum}
  \Delta_{\mathcal{R}}^2 = \left(\frac{\D\langle\phi\rangle}{H\D t}\right)^{-2} \frac{\D \text{var}\!\left(\phi\right)}{H\D t}\,.
\end{equation}
The calculation of the scalar power spectrum -- as was previously found in~\cite{Feldstein:2006hm} -- reduces to the determination of the mean inflaton field value and its variance. 

\begin{figure}[t!]
\begin{center}
  \includegraphics[width=5cm]{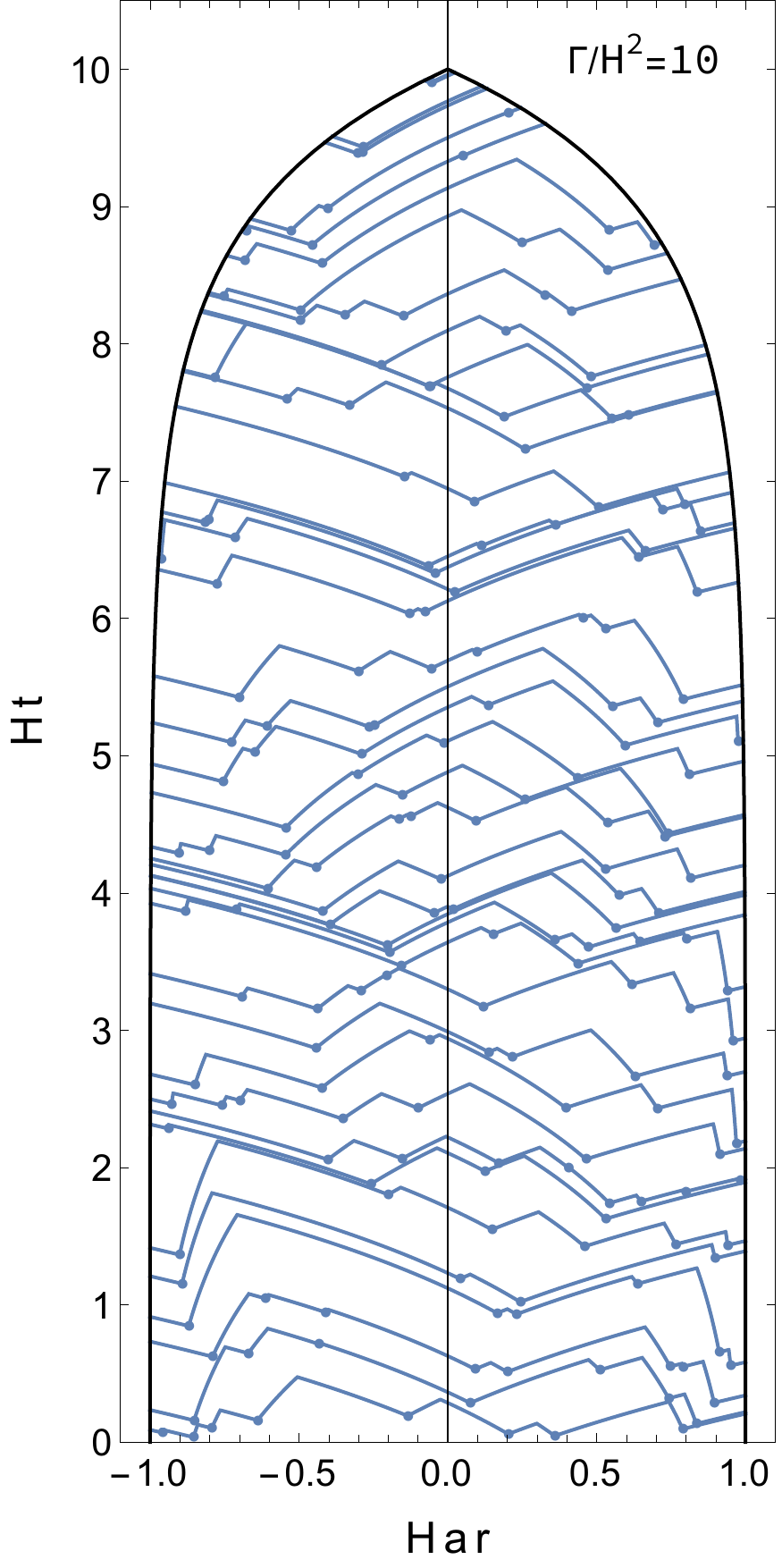}$\;$
  \includegraphics[width=5cm]{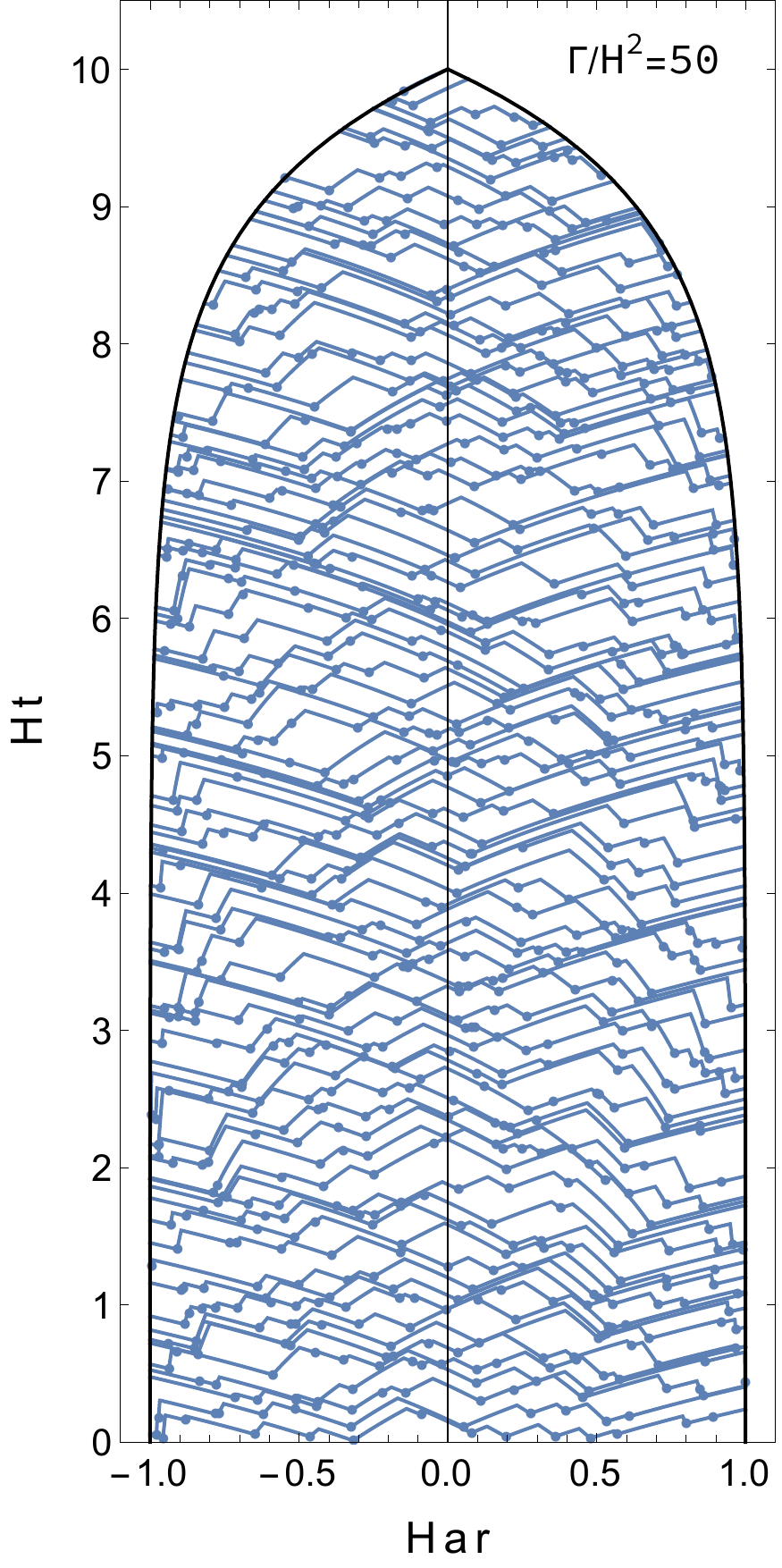}$\;$
  \includegraphics[width=5cm]{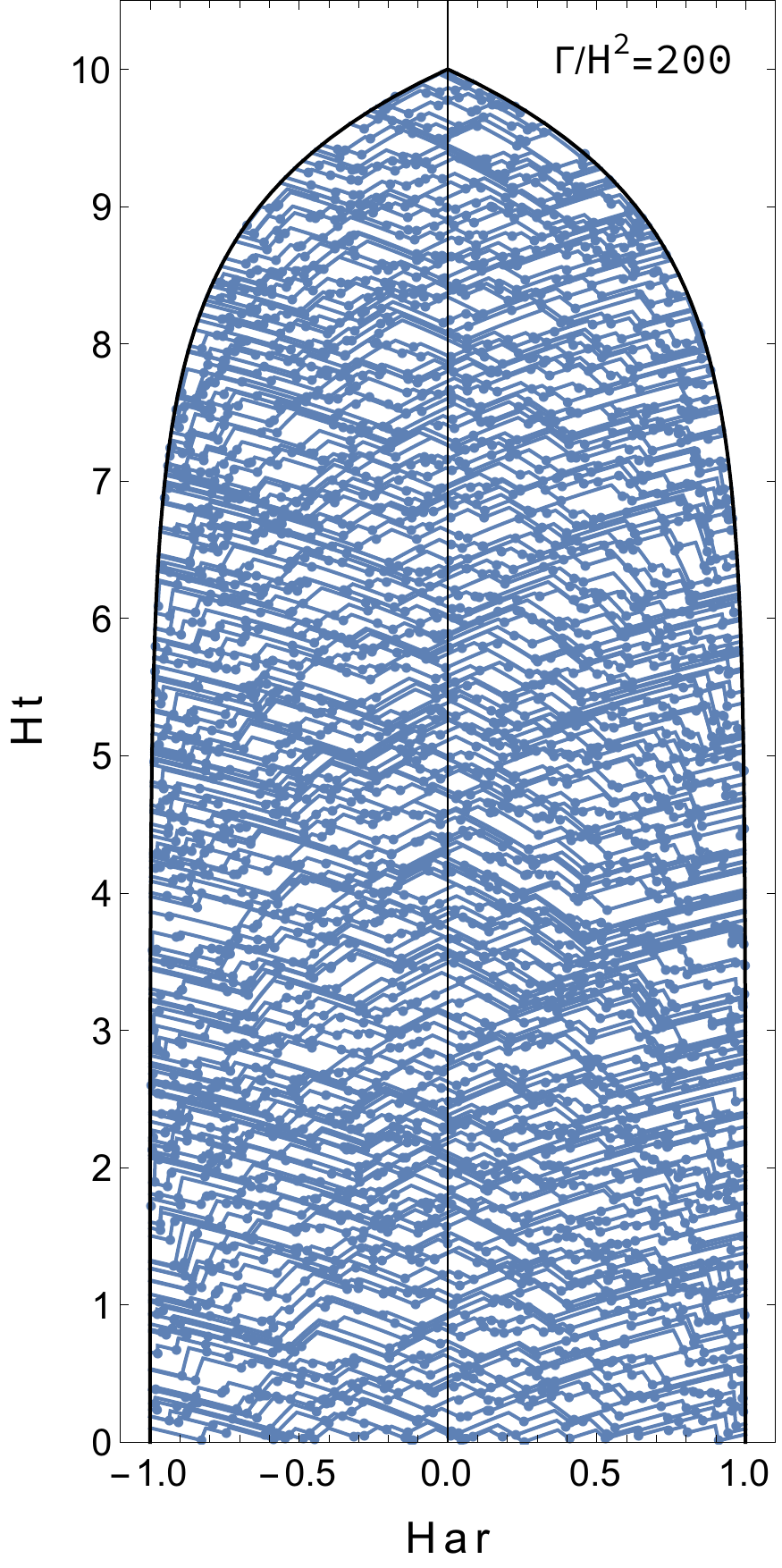}
\end{center}
\vspace{-6mm}
\caption{Evolution of bubble walls in a past light cone diagram in 1+1 dimensions. Note that time moves upward in the diagram; the spacetime point of interest is at the top of the diagram and looking at the
past light cone means looking downward from this point. In the three panels the tunneling rate increases from left to right. The horizontal axis denotes the position in real space (not the comoving position) in Hubble units. Vacuum transitions are depicted by points, while the lines correspond to bubble walls. The total number of vacuum transitions $N_W$ is obtained by counting the number of walls hitting an observer at $\position=0$. The three panels above feature $N_W=\{40, 90, 182\}$ (from left to right).} 
\label{fig:lightcone}
\end{figure}

For illustration, it is instructive to look at the problem of finding $\langle\phi\rangle$ and $\text{var}\phi$ in 1+1 dimensions. In Fig.~\ref{fig:lightcone} we draw the past light cone of the space-time point $(\position,t)=(0,10\,H^{-1})$ in real space. Note that time moves upward in the diagram; the spacetime point of interest is at the top of the diagram and looking at the
past light cone means looking downward from this point. Due to the scale-invariance of the power spectrum we could have selected essentially any other space-time point for the evaluation. The only requirement is that, in order to ensure that initial conditions are sufficiently `washed out', the time should be chosen at least a few e-folds after the beginning of inflation (the latter corresponding to $t=0$). We randomly distributed $N$ tunneling events in the past light cone which are depicted as points in the figure. The number $N$ follows a Poisson distribution with mean value
\begin{equation}\label{eq:Nmean}
  \langle N \rangle = \mathcal{V}_2\, \Gamma\,,
\end{equation}
where the two-volume enclosed by the past light cone is denoted by $\mathcal{V}_2$. The three panels in Fig.~\ref{fig:lightcone} reflect different choices of $\Gamma$. Each transition creates a bubble containing the next vacuum in the chain. The bubble walls (depicted as lines in the figure) expand approximately at the speed of light. Whenever two bubble walls collide, they disappear by producing radiation which quickly redshifts away.

In 3+1 dimensions a completely analogous picture arises, we just need to replace the two-volume $\mathcal{V}_2$ by the four-volume $\mathcal{V}_4$ in~\eqref{eq:Nmean}. For the determination of $d\phi/dt$, we need to count the number of bubble walls $N_W$ hitting an observer at $\position=0$ within the past light cone. Due to the linear time evolution, we then have 
\begin{equation}\label{eq:dphi}
\frac{\D\phi}{H \D t} = \frac{N_W}{Ht}\times\Delta\phi\,,
\end{equation}
where $\Delta\phi$ denotes the field space distance between two neighboring minima in the potential. By randomly generating a large number of realizations, we can finally determine $\D\langle \phi\rangle/\D t$ and $\D\text{var}\phi / \D t$.

\subsection{Results of Simulations}

\begin{figure}[h]
\begin{center}
  \includegraphics[height=7cm]{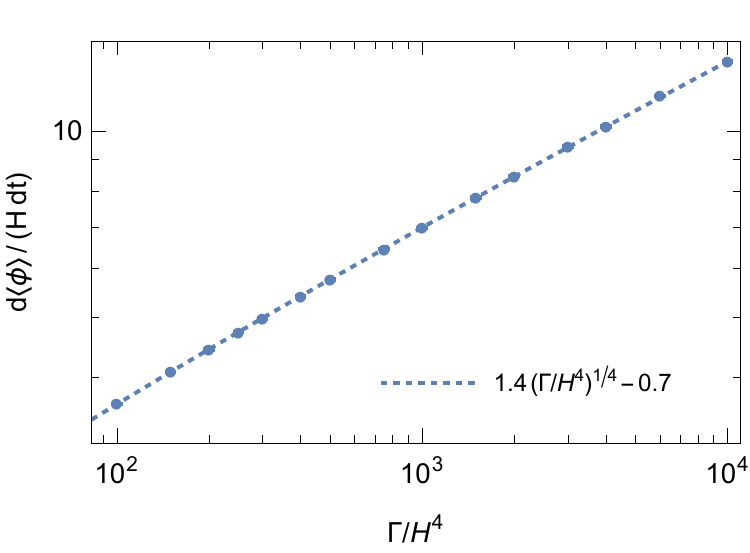}
\end{center}
\vspace{-6mm}
\caption{Derivative of the mean inflaton field value from our simulation (error bars). Statistical errors are tiny and, therefore, hardly visible in the figure. The dashed line indicates the fit shown in the plot legend. The mean inflaton field value scales linearly with $\Delta\phi$ which we set to unity in the figure (to avoid clutter).} 
\label{fig:simulations}
\end{figure}

Before we discuss the results of our simulations in 3+1 dimensions, we can try to anticipate the dependence of $\D\langle \phi\rangle/\D t$ on the tunneling rate. We first note that the mean free path of a bubble wall\footnote{We define the mean free path of a bubble wall as the average distance it has traveled upon collision with another bubble wall.} roughly corresponds to the mean space-time distance between two tunneling events $\Gamma^{-1/4}$. Only bubble walls originating from within a radius $\sim \Gamma^{-1/4}$ typically reach an observer at $\position=0$, while walls from more distant tunneling events tend to be erased by collisions with other walls before making it to $\position=0$. The fraction $N_W$ of the total number $N$ of tunneling events in the past light cone which trigger an actual transition can, hence, be estimated as $\langle N_W \rangle /\langle N \rangle \sim (\Gamma^{-1/4} H)^3$ (see~\cite{Bedroya:2020rac} for a similar argument). According to~\eqref{eq:dphi} we then have
\begin{equation}\label{eq:dphi4d}
\frac{\D\langle\phi\rangle}{H \D t} \sim \left(\frac{\Gamma}{H^4}\right)^{1/4} \times\Delta\phi\,.
\end{equation}
We have run excessive simulations in order to verify the above scaling. For values of $\Gamma/H^4=10^2 - 10^4$, we randomly generated tunneling patterns in the past light cone and extracted the number of vacuum transitions for a large number of independent realizations.\footnote{Since the computation time rapidly increases with the number of tunneling events in the past light cone, the number of realizations in our simulations decreases from $\sim 10^5$ at $\Gamma/H^4=10^2$ to $\sim 10^3$ at $\Gamma/H^4=10^4$.} From the mean number of transitions and its variance, we derived $\D \langle \phi \rangle/ \D t$ and $\D \text{var}\phi/ \D t$ and the corresponding statistical error.\footnote{In order to derive the statistical error on the variance we also extracted the fourth moment of the inflaton field value from our simulations.} In Fig.~\ref{fig:simulations} we depict the results for the mean inflaton field value. As can be seen, a good fit to the simulations is obtained with the function
\begin{equation}\label{eq:dphi4dsim}
\frac{\D\langle\phi\rangle}{H \D t} \simeq \left(1.4\, \left(\frac{\Gamma}{H^4}\right)^{1/4} -0.7\right) \Delta\phi\,,
\end{equation}
in good agreement with the expectation~\eqref{eq:dphi4d}. Since a fit with free power law index yields 0.257 as its best fit value, we have set the index to the physically motivated value of 1/4 above. For a small number of tunneling events slight deviations from the power law behavior occur which manifest as the offset by 0.7 in~\eqref{eq:dphi4dsim}. However, observations will turn out to favor large values of $\Gamma/H^4$ such that we can neglect the offset in the following.

\begin{figure}[h]
\begin{center}
  \includegraphics[height=7cm]{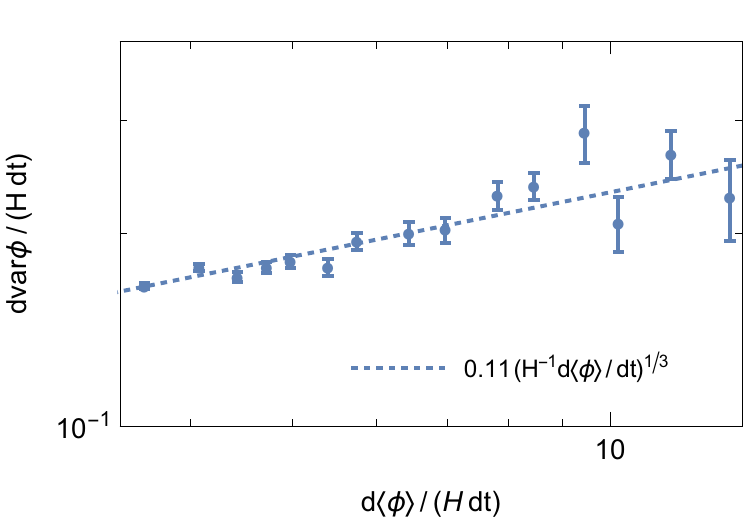}
\end{center}
\vspace{-6mm}
\caption{Derivative of the variance from our simulation (error bars) expressed by the derivative of the mean inflaton field value. The variance scales quadratically with $\Delta\phi$ which we set to unity in the figure.} 
\label{fig:simulations2}
\end{figure}

In the absence of an analytic estimate we have to rely on our simulations to determine the inflaton variance. Fig.~\ref{fig:simulations2} depicts the variance in terms of the mean field value. We again observe a power law behavior which is well described by the fit
\begin{equation}\label{eq:dvarphi4dsim}
\frac{\D\text{var}\phi}{H \D t} \simeq 0.11 \left(\frac{\D\langle\phi\rangle}{\Delta\phi H \D t}\right)^{1/3} \left(\Delta\phi\right)^2\,.
\end{equation}
Since our numerical fit yielded 0.33 as best fit power law index we anticipated that the exact index in the expression above is 1/3. 

Inserting~\eqref{eq:dphi4dsim} and~\eqref{eq:dvarphi4dsim} into~\eqref{eq:powerspectrum} we finally obtain the scalar power spectrum of chain inflation
\begin{equation}\label{eq:powerfinal}
\Delta_{\mathcal{R}}^2 \simeq 0.06 \left(\frac{\Gamma}{H^4} \right)^{-5/12}\,.
\end{equation}
Notice that $\Delta\phi$ cancels out in the above expression. The power spectrum only depends on the tunneling rate and the Hubble rate\footnote{Although there is no explicit dependence on $\Delta\phi$, it does play a role in the tunneling rate.}. We emphasize that our result~\eqref{eq:powerfinal} is in agreement with the the calculation of Feldstein \& Tweedie~\cite{Feldstein:2006hm} (cf.~\eqref{eq:powerprevious}).

At this point we should make a further remark: our derivation of the power spectrum relies on the assumption that colliding bubble walls quickly transfer their energy into radiation which redshifts away. Under certain circumstances wall collisions can, however, also trigger new vacuum transition~\cite{Easther:2009ft,Giblin:2010bd}. In fact, the power spectrum calculation of Cline et al.~\cite{Cline:2011fi} is based on this alternative assumption. While~\eqref{eq:powerspectrum} remains valid in this case, the mean inflaton field value would simply be given by the number of tunneling processes in the past lightcone. This is because every bubble wall at some point would collide with another bubble wall and trigger the next transition in the chain such that $\langle \phi(t) \rangle = \Gamma \mathcal{V}_4 \Delta\phi \simeq 4\pi \Gamma t \Delta\phi/(3\, H^3)$. Since the total number of tunneling events follows Poissonian statistics $\text{var}\phi(t)/ \Delta\phi= \langle \phi(t) \rangle $ in this case. With the help of~\eqref{eq:powerspectrum} one then immediately obtains the power spectrum of Cline et al.~\cite{Cline:2011fi} shown in~\eqref{eq:powerprevious} without the need of simulations. 

However, we note that the assumption that bubble wall collisions trigger new transitions requires that the dominant fraction of the wall energy is transferred to kinetic energy of the inflaton. This can easily be motivated in simple one-field models. On the other hand, in more complete particle realizations with further degrees of freedom to which the inflaton couples, additional dissipation mechanisms exist and typically dominate. In this light, we will focus on the case, where colliding bubble walls transfer their energy into radiation which quickly disappears. Under this assumption,~\eqref{eq:powerfinal} is the correct scalar power spectrum. 

Let us finally remark that yet other calculations~\cite{Watson:2006px,Huang:2007ek,Chialva:2008zw,Chialva:2008xh} listed in~\eqref{eq:powerprevious} obtained a power spectrum resembling the one from slow roll inflation.
We believe that these spectra do not apply to chain inflation, where the origin of fluctuations is different compared to the slow roll scenario.

\subsection{Comparison with CMB Observables}\label{sec:cmbobservables}

In the first step, we constrain chain inflation by the measured amplitude of the scalar power spectrum~\cite{Bunn:1996py,Akrami:2018odb} 
\begin{equation}
A_s \equiv \Delta_{\mathcal{R}}^2|_{k=k_*}=2.1\cdot 10^{-9}\,,
\end{equation} 
where $k_*$ denotes the pivot scale. Comparison with~\eqref{eq:powerfinal} reveals
\begin{equation}\label{eq:rateconstraint}
\frac{\Gamma}{H^4} \simeq 8\times 10^{17}\,.
\end{equation}
Viable chain inflation thus requires a rather large tunneling rate. We remind the reader that our calculation did not include the fluctuations sourced by bubble wall collisions which could add another contribution to the power spectrum. Conservatively, one may, hence, interpret~\eqref{eq:rateconstraint} as a lower limit on the tunneling rate.

The number $\Gamma/H^4$ measures the tunneling events per Hubble volume. In order to translate this to the number of transitions  $N_t$ that the inflaton undergoes per e-folding, we calculate
\begin{equation}\label{eq:transitionnumber}
N_t = \left.\frac{1}{\Delta\phi}\frac{d\langle \phi \rangle}{H\D t}\right|_{k=k_\star} \simeq 4.2\times 10^{4}\, ,
\end{equation}
where we used~\eqref{eq:dphi4dsim} and took into account that the field value changes linearly with time.\footnote{This is again in reasonable agreement with the number of $\sim 3.5\times 10^{4}$ transitions quoted in~\cite{Feldstein:2006hm}.} Notice that the number of transitions is much smaller than the number of tunneling events per Hubble volume. As explained previously this is because a tunneling event only triggers a new transition at some spatial location $\position$ if its bubble wall reaches $\position$ before colliding with other walls (which becomes highly unlikely if the tunneling event occurs at a distance further than $\Gamma^{-1/4}$ from $\position$, see Fig.~\ref{fig:lightcone} for illustration).

So far we have treated $\Gamma$, $H$ as constants. In the more realistic case, where both rates vary slowly with time, we expect~\eqref{eq:powerfinal} to remain approximately valid. However, the time-dependence of $\Gamma$, $H$ introduces a (slight) derivation from scale-invariance in the power spectrum. We can estimate the scalar spectral index as
\begin{equation}\label{eq:ns}
  n_s = 1+\left.\frac{\D \log \Delta_{\mathcal{R}}^2}{\D \log k}\right|_{k=k_*} \simeq 1+ \frac{5}{12}\,\left( \frac{4\dot{H}}{H^2} - \frac{\dot{\Gamma}}{H \Gamma} \right)\,,
\end{equation}
where we approximated $\dot{k}\simeq \dot{a}H=aH^2$. All quantities on the right side must be evaluated at the pivot scale relevant for CMB observations. 

Notice that we can relate $\dot{H}$ to the tunneling rate
\begin{equation}\label{eq:Hdot}
\frac{\dot{H}}{H^2} = -\frac{\Delta H}{H }\frac{\D \langle\phi\rangle}{\Delta \phi \,H\, \D t} \simeq -1.4\,\frac{\Delta H}{H }\left(\frac{\Gamma}{H^4}\right)^{1/4} \simeq -4\times 10^{4}\,\frac{\Delta H}{H }\,,
\end{equation}
where we employed~\eqref{eq:dphi4dsim} and imposed the normalization of the power spectrum~\eqref{eq:rateconstraint} in the last step. The (positive) parameter $\Delta H$ denotes the difference in Hubble scale between two neighboring minima. Since the Hubble scale necessarily decreases with time, a red spectral index as required by data ($n_s = 0.965$~\cite{Akrami:2018odb}) can be realized with $\Delta H/H \sim 5\times 10^{-7}$ and vanishing $\dot{\Gamma}$. However, in general, we expect contributions to $n_s$ from both, $\dot{H}$ and $\dot{\Gamma}$.

\subsection{Swampland Conjecture}\label{sec:swampland}
It was recently proposed that any field theory consistent with quantum gravity should fulfill the Trans-Planckian Censorship Conjecture~\cite{Bedroya:2019snp}. The latter requires the life-time $\tau$ of a metastable de Sitter space to fulfills the constraint
\begin{equation}
H \tau < \log H^{-1}\,,
\end{equation}
in Planck units for (approximately) constant $H$. In chain inflation $H \tau$ is simply the inverse of $N_t$ given in~\eqref{eq:transitionnumber}. The correct normalization of the scalar power spectrum thus requires
\begin{equation}
H\tau = 2.4 \times 10^{-5}\,.
\end{equation}
We observe that the Trans-Planckian Censorship Conjecture is automatically satisfied in viable models of chain inflation.\footnote{We note that, since $\tau\simeq \Gamma^{-1/4}$, the Trans-Planckian Censorship Conjecture implies $\Gamma^{1/4}/H>1$ up to $\mathcal{O}(1)$ numbers. This was also found  in~\cite{Bedroya:2020rac} and denoted as the membrane version of the conjecture.}

\section{Tunneling Rates in Viable Models of Chain Inflation}

In the previous section, chain inflation has been constrained by the observables of the cosmic microwave background. Now we turn to the implications for model realizations.

\subsection{Tunneling Formalism}

For this purpose, we take a closer look at the tunneling rate of a scalar field $\phi$ in a potential $V$ with Lagrangian
\begin{equation}
  \mathcal{L}= \frac{1}{2}\partial_\mu \phi \partial^\mu \phi - V(\phi)\,.
\end{equation}
We note that chain inflation can operate in wide class of theories including realizations in the string landscape~\cite{Freese:2004vs,Freese:2006fk,Chialva:2008zw,Chialva:2008xh}, field theoretic axion models~\cite{Freese:2005kt,Ashoorioon:2008pj} and even scenarios without fundamental scalars~\cite{Watson:2006px}. In many instances, however, an effective description in terms of a single scalar field applies to the tunneling processes (while a more complete picture of the interaction part may be required to trace e.g.\ bubble wall collisions).

The tunneling rate (per unit four volume) between two minima of the potential takes the form
\begin{equation}\label{eq:gamma}
  \Gamma = A \, e^{-S_E}\,,
\end{equation}
where $S_E$ denotes the Euclidean action of the bounce solution extrapolating between the two vacua. The bounce solution is obtained by solving the differential equation
\begin{equation}\label{eq:diffeq}
  \frac{\D^2\phi}{\D\rho^2} + \frac{3}{\rho}\frac{\D\phi}{\D\rho} = V'(\phi)\,,
\end{equation}
with the boundary condition
\begin{equation}
  \lim\limits_{\rho\rightarrow\infty} \phi(\rho) = \phi_+\,,
\end{equation}
where $\phi_+$ indicates the minimum in which $\phi$ is initially located. The commonly applied thin-wall approximation~\cite{Coleman:1977py} requires a small energy difference between the two vacua participating in the tunneling. However, individual vacua tend to be extremely long-lived in this case. In~\cite{Freese:2005kt} it was shown that proper percolation and thermalization restricts chain inflation to the regime outside the validity of the thin-wall approximation. A more accurate estimate of tunneling rates in chain inflation has been obtained in~\cite{Ashoorioon:2008pj}. The latter is based on a triangle approximation of the potential between two minima~\cite{Duncan:1992ai}.

In the following we will further improve the calculation of tunneling rates in chain inflation by exactly solving the differential equation of the bounce. In contrast to previous work we will also explicitly address the tunneling prefactor $A$ in~\eqref{eq:gamma}.

\subsection{Exact Tunneling Rates in a Periodic Potential}

\begin{figure}[h!]
\begin{center}
  \includegraphics[width=8cm]{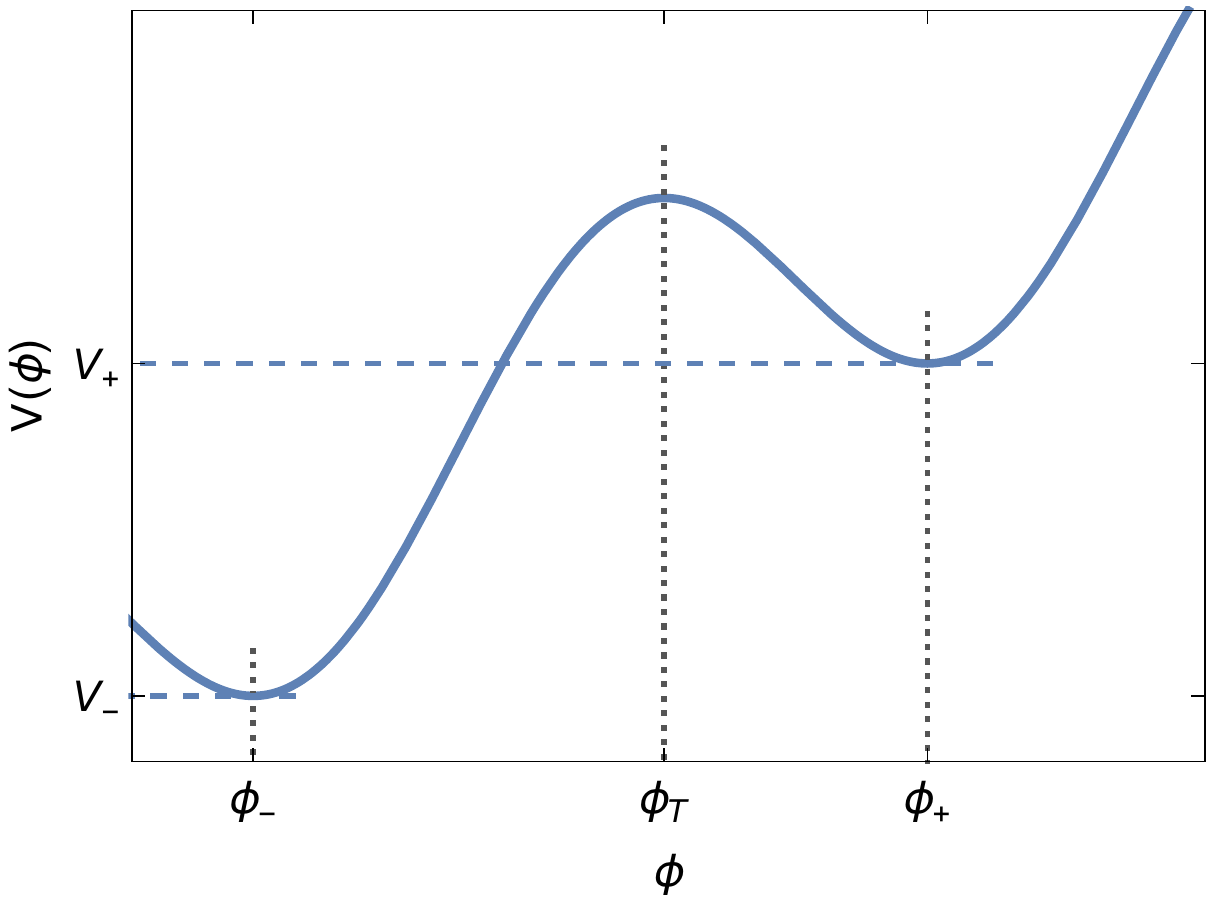}
\end{center}
\vspace{-6mm}
\caption{Typical potential in chain inflation between two minima ($\phi_-$,$V_-$) and ($\phi_+$,$V_+$). The field-value of the intermediate maximum is denoted by $\phi_T$.}
\label{fig:potential}
\end{figure}

We will assume that locally the potential of the inflaton can be approximated as
\begin{equation}\label{eq:potential}
  V(\phi)=\mu^3\phi + \Lambda^4 \cos\frac{\phi}{f} + \text{const} \,.
\end{equation}
For the moment, we shall assume that gravitational corrections are negligible such that tunneling is not affected by the constant term (below we will confirm that these corrections are indeed negligible). Fig.~\ref{fig:potential} depicts the potential between two minima.

The potential~\eqref{eq:potential} was previously studied in~\cite{Freese:2005kt} on chain inflation with axions (see also~\cite{Abbott:1984qf}). It is also motivated by models of axion monodromy~\cite{Silverstein:2008sg,McAllister:2008hb,Flauger:2009ab}, modulated natural inflation~\cite{Kim:2004rp,Kappl:2015esy,Winkler:2019hkh} and winding inflation~\cite{Hebecker:2015rya,Hebecker:2018fln}, in which a discrete axion shift symmetry is weakly broken (either by fluxes or by the interplay with a second axion). We do, in general, not require~\eqref{eq:potential} to hold globally, but to yield a good approximation between two neighboring minima. In the mentioned schemes $\Lambda$, $\mu$ and (possibly) $f$ are field-dependent, but they typically vary on scales much larger than the distance $2\pi f$ between two minima such that they can locally be approximated as constants. 

Besides the theoretical motivation, we can in fact approximate an arbitrary potential with two minima by~\eqref{eq:potential} through the following map
\begin{equation}\label{eq:map}
   \mu^3 = \frac{V_+ - V_-}{\phi_+ - \phi_-}\,,\qquad
  f =\frac{\phi_+ - \phi_-}{2\pi}\,,\qquad
  \Lambda^4=\frac{V_+ - V_-}{2\pi\sin\left[\pi \left( \frac{1}{2}- \frac{\phi_+ - \phi_T}{\phi_+ - \phi_-} \right) \right]} \,,
\end{equation}
where $\phi_{\pm}$ and $V{\pm}$ denote field value and potential at the higher and lower minimum, and $\phi_T$ the field value of the maximum in between (see Fig.~\ref{fig:potential}). In the following, we derive the tunneling rate between two successive minima as a function of $\Lambda$, $\mu$ and $f$. 

\begin{figure}[h!]
\begin{center}
  \includegraphics[width=10.6cm]{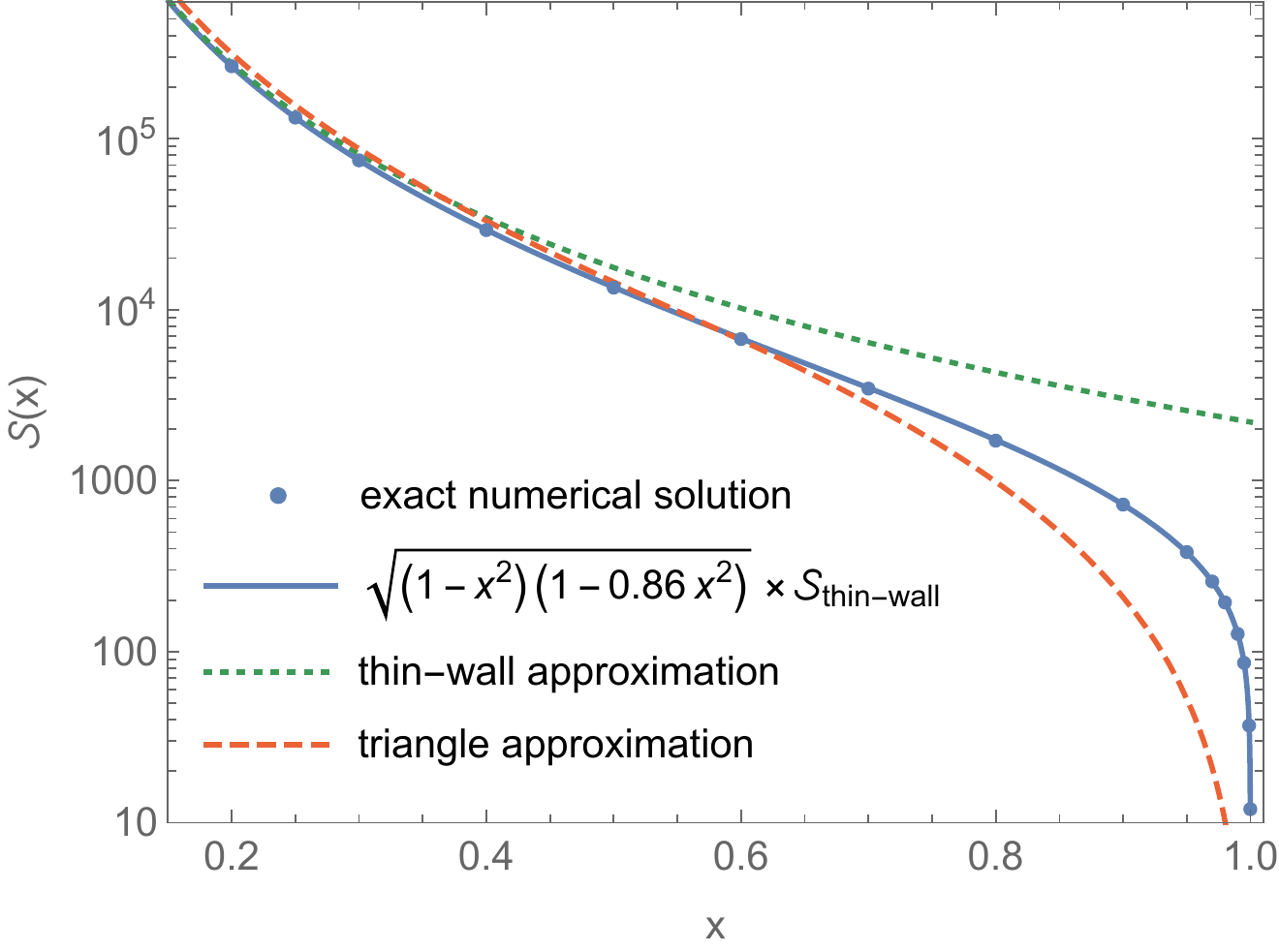}
\end{center}
\vspace{-6mm}
\caption{The rescaled bounce action $\mathcal{S}$ (defined in~\eqref{eq:s}) as a function of $x=f\mu^3/\Lambda^4$. The exact numerical solution evaluated at several $x$ (points) is shown together with the analytic fit indicated in the plot legend. Also shown is $\mathcal{S}$ as obtained in the thin-wall and the triangle approximation.}
\label{fig:bounceaction}
\end{figure}

The calculation can be drastically simplified by employing the symmetries of~\eqref{eq:diffeq}. In this way, one finds that the bounce action for the potential~\eqref{eq:potential} can be expressed as
\begin{equation}\label{eq:s}
  S_E = \frac{f^4}{\Lambda^4} \;\mathcal{S}\!\left(\frac{f\mu^3}{\Lambda^4}\right)\,.
\end{equation}
Here, we introduced the rescaled bounce action $\mathcal{S}$ which only depends on the combination $x = f\mu^3/\Lambda^4$. In the thin-wall approximation $\mathcal{S}$ can be determined analytically and one obtains~\cite{Freese:2005kt}
\begin{equation}\label{eq:sthinwall}
 \mathcal{S}_{\text{thin-wall}}(x) = \frac{4}{\pi} \left(\frac{12}{x}\right)^3\,.
\end{equation}
In Fig.~\ref{fig:bounceaction} we compare the thin-wall approximation to the exact bounce action obtained by a numerical solution of~\eqref{eq:diffeq}. It can be seen that the thin-wall approximation is applicable in the regime of $x\lesssim 0.4$. The triangle approximation~\cite{Duncan:1992ai,Ashoorioon:2008pj} also shown in Fig.~\ref{fig:bounceaction} can be used up to $x\sim 0.7$. However, in the regime of fast tunneling ($x\gtrsim 0.8$) which is favored by chain inflation (cf.~\eqref{eq:rateconstraint}), both aproximations deviate significantly from the exact numerical result. We, therefore, now provide a new analytic approximation of the bounce action which was obtained by a fit to the numerical result,
\begin{equation}\label{intermediatewall}
\mathcal{S}(x) = \sqrt{(1-x^2)\,(1-0.86 x^2)}\times \mathcal{S}_{\text{thin-wall}}(x)\,,
\end{equation}
with $\mathcal{S}_{\text{thin-wall}}$ from~\eqref{eq:sthinwall}. The fit is also depicted in Fig.~\ref{fig:bounceaction}. 

The pre-exponential factor in the tunneling rate~\eqref{eq:gamma} is calculated by considering quantum fluctuations about the classical action of the bounce. It can be expressed in terms of functional determinants,
\begin{equation}\label{eq:prefactor}
A = \frac{S_E^2}{4\pi^2} \left|\frac{\det^\prime (-\partial^2+ V^{\prime\prime}(\phi_B))}{\det (-\partial^2+ V^{\prime\prime}(\phi_+))} \right|^{-1/2}\,,
\end{equation}
where $\det^\prime$ denotes the determinant with the zero modes removed, while $\phi_B$ stands for the bounce solution for the field.\footnote{Since~\eqref{eq:prefactor} is a one-loop expression, divergences must be absorbed by also adding a counterterm in the exponent of~\eqref{eq:gamma}.}

In order to avoid a tedious numerical evaluation of the determinants, we want to obtain an analytic approximation in terms of the potential parameters. In~\cite{Baacke:2003uw} the leading contribution to the prefactor has been found to derive from the Born approximation term $\mathcal{A}^{(1)}_{\text{fin}}$ (see also~\cite{Dunne:2005rt}),
\begin{equation}\label{eq:A1fin}
A \simeq m^4\,\frac{S_E^2}{4\pi^2}\:\exp\left(-\frac{\mathcal{A}^{(1)}_{\text{fin}}}{2}\right)\,,\qquad
\mathcal{A}^{(1)}_{\text{fin}} = 
-\frac{m^2}{8}\int\limits_0^\infty \D \rho\, \rho^3 \,\left(V''(\phi_B(\rho))-m^2\right)
\,,
\end{equation}
where we introduced the mass term $m^2 = V''(\phi_+)$. The approximation is most accurate in the regime of fast tunneling where the barrier height between minima gets small~\cite{Baacke:2003uw}.

\begin{figure}[h!]
\begin{center}
  \includegraphics[width=10.6cm]{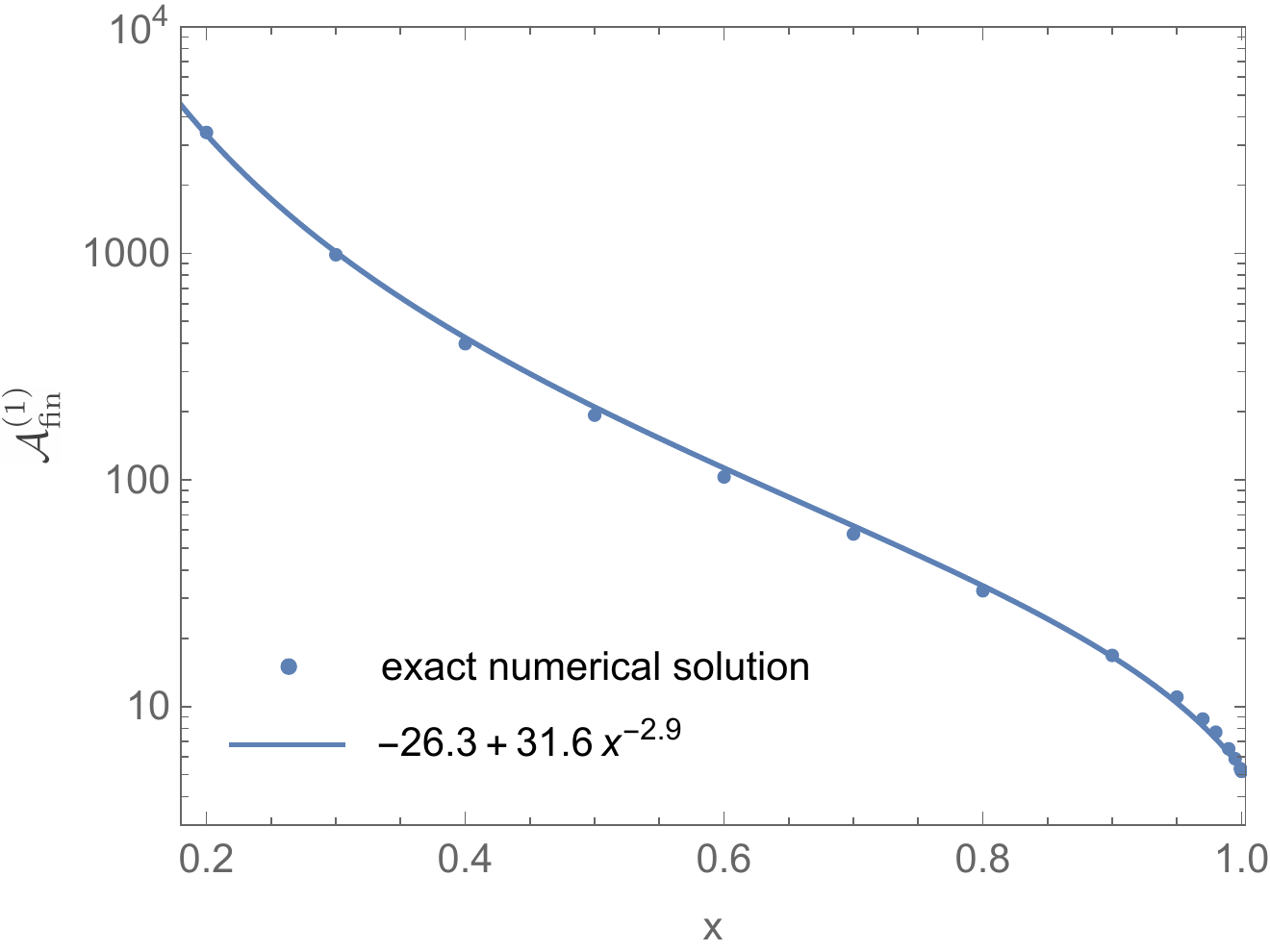}
  \end{center}
\vspace{-6mm}
\caption{The Born approximation term (defined in~\eqref{eq:A1fin}) as a function of $x=f\mu^3/\Lambda^4$. The exact numerical solution evaluated at several $x$ (points) is shown together with the analytic fit indicated in the plot legend.}
\label{fig:prefactor}
\end{figure}

For the potential considered here, it can be shown that $\mathcal{A}^{(1)}_{\text{fin}}$ only depends on the parameter combination $x=f\mu^3/\Lambda^4$. The result of a numerical evaluation of $\mathcal{A}^{(1)}_{\text{fin}}$ in terms of $x$ is depicted in Fig.~\ref{fig:prefactor}. The numerical solution is well-approximated by the fit function
\begin{equation}
\mathcal{A}^{(1)}_{\text{fin}}(x) \simeq -26.3+31.6 x^{-2.9}\,,
\end{equation}
as shown in the same figure. 

Expressing $m^4=(\Lambda^8/f^4)(1-x^2)$, our final approximation for the tunneling rate in the potential~\eqref{eq:potential} reads
\begin{equation}\label{eq:final_approximation}
\Gamma = \frac{\Lambda^8}{f^4}\,(1-x^2)\,\frac{S_E^2}{4\pi^2}\:\exp\left(13.15 - \frac{15.8}{x^{2.9}}\right)
\times \exp\left(-S_E\right)
\end{equation}
with
\begin{equation}
S_E=\frac{f^4}{\Lambda^4}\sqrt{(1-x^2)\,(1-0.86 x^2)}\;\frac{4}{\pi} \left(\frac{12}{x}\right)^3\,,\qquad x = \frac{f\mu^3}{\Lambda^4}\,.
\end{equation}
While our result for the bounce action directly follows from the standard bounce formalism, several approaches to address the functional determinants in the tunneling prefactor have been discussed in the literature. Our approximation for the tunneling prefactor relies on the validity of the formalism described in~\cite{Baacke:2003uw}.

\subsection{Implications of CMB Constraints for Chain Inflation with Axions}

\begin{figure}[h!]
\begin{center}
  \includegraphics[width=10.6cm]{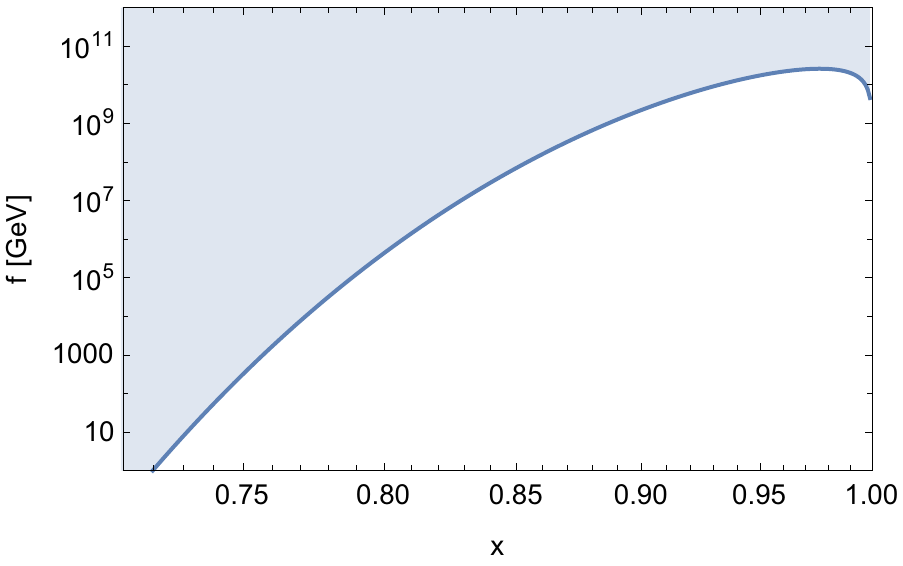}
\end{center}
\vspace{-6mm}
\caption{Constraint on the axion decay constant as a function of $x=f\mu^3/\Lambda^4$. The blue shaded region is excluded by the combination of CMB and unitarity constraints.}
\label{fig:fmax}
\end{figure}

We can now constrain axion models of chain inflation by imposing that their scalar power spectrum matches with observations. The requirement of a nearly scale-independent spectrum implies\footnote{We assume absence of excessive fine-tuning between $\dot{H}$ and $\dot{\Gamma}$ in~\eqref{eq:ns}. It appears very unlikely that such fine-tuning can be realized over the entire range of scales accessible through CMB observations.} (cf.~\eqref{eq:ns})
\begin{equation}
\frac{|\dot{H}|}{H^2} \simeq 4\times 10^{4}\,\frac{\Delta H}{H } \lesssim 1\,.
\end{equation} 
For the model under consideration we can set $\Delta H \sim f \mu^3/H $, so that the requirement of approximate scale invariance becomes
\begin{equation}\label{eq:Hubble}
H^2 \gtrsim 4\times 10^{4}\,f \mu^3
\end{equation}
in Plank units. As an additional constraint we require the theory~\eqref{eq:potential} to remain in the perturbative regime. This imposes a perturbative unitarity limit on the quartic coupling\footnote{A similar constraint has also been applied to chain inflation in~\cite{Cline:2011fi}.}
\begin{equation}\label{eq:unitarity}
\left.\frac{\D^4 V }{\D \phi^4}\right|_{\phi=\phi_{\pm}}
= \frac{\Lambda^4}{f^4}\,\sqrt{1-x^2} < 8\,\pi\,.
\end{equation}
Combining~\eqref{eq:Hubble} and~\eqref{eq:unitarity} with~\eqref{eq:final_approximation}, we obtain the following constraint on the tunneling rate per Hubble four-volume,
\begin{equation}\label{eq:maxgamma}
\frac{\Gamma}{H^4} < \frac{0.06\,(1-x^2)^3\,(1-0.86\,x^2)}{f^4\,x^8}\times\exp\left(-\frac{(1-x^2)\sqrt{1-0.86\,x^2}}{2\pi^2}\left(\frac{12}{x}\right)^3   - \frac{15.8}{x^{2.9}}\right)\,.
\end{equation}
Since the amplitude of scalar perturbations fixes the tunneling rate to $\Gamma/H^4 \simeq 8\times 10^{17}$ (see Sec.~\ref{sec:cmbobservables}), the above expression translates to a strong upper limit on the axion decay constant. The maximal $f$ consistent with the power spectrum normalization is depicted in Fig.\ref{fig:fmax}. For any value of $x$, the axion decay constant must satisfy
\begin{equation}\label{eq:fmax}
f < 3 \times 10^{10}\gev\,.
\end{equation}

A word of caution is warranted if $x$ approaches unity as the minima in the potential become shallow. In this regime the inflaton may trigger a tunneling catastrophe, where it overshoots the next minimum in the potential and directly tunnels to the bottom of the potential. If we translate the constraint from~\cite{Cline:2011fi} to the model under consideration, the regime with $x>0.96$ is disfavored. While additional dissipation mechanisms may weaken the bound on $x$, even taken at face value it would only marginally affect the limit on $f$ (see Fig.~\ref{fig:fmax}).

A more significant impact on the axion decay constant can occur if the potential deviates from the assumed cosine-shape. Unless for extreme choices, we do, however still expect~\eqref{eq:fmax} to at least approximately hold.

The upper limit on the axion decay constant may impose a model building challenge for the realization of chain inflation in string theory, where axion decay constants typically come out not too far below the Planck scale. On the other hand, several mechanisms to suppress the decay constants of string axions have been discussed (see e.g.~\cite{Svrcek:2006yi}). In fact, $f\sim 10^{9}-10^{12}\gev$ is favored in the axion solution of the strong CP problem. While the axion driving chain inflation cannot be identified with the QCD axion -- the small QCD scale in relation to the axion decay constant would suppress the tunneling rate too much~\cite{Ashoorioon:2008pj} -- model building challenges related to a small axion decay constant can be addressed in a similar way. 

There is another motivation to consider axion decay constants significantly below the Planck scale. The smallness of f is not only required by the CMB constraints, it also ensures that the tunneling rate satisfies the Swampland conjecture on Trans-Planckian Censorship (see Sec.~\ref{sec:swampland}). In this light, $f\ll M_P$ could even turn out favorable for the implementation of chain inflation into consistent ultraviolet theories.

Let us finally comment on gravitational effects on the tunneling rate which we neglected so far. For this purpose we need to estimate the radius $R$ of the nucleated bubbles which is given by the bubble tension divided by potential difference $\Delta V = V_+ -V_-$ between minima. Extracting the tension from~\cite{Freese:2005kt} and neglecting $\mathcal{O}(1)$ numbers we obtain
\begin{equation}
R \sim \frac{\Lambda^2}{\mu^3} \sim \frac{f}{\Lambda^2}\,,
\end{equation}
where we employed that $x=f\mu^3/\Lambda^4$ cannot be too far from unity. Otherwise the tunneling rate would be so highly suppressed that the power spectrum constraint~\eqref{eq:rateconstraint} cannot be satisfied.

We can also use that even for $S_E\sim 1$ the tunneling rate~\eqref{eq:final_approximation} is bounded from above by the tunneling prefactor and, hence, $\Gamma \lesssim \Lambda^8/f^4$. This implies
\begin{equation}
R \lesssim \Gamma^{-1/4} \ll H^{-1}\,
\end{equation}
where we again imposed the normalization of the scalar power spectrum in the last step. In the parameter regime of viable chain inflation, which we identified in this work, the bubble radius is much smaller than the de Sitter horizon. This justifies our omission of gravitational corrections~\cite{Ibanez:2015fcv}. 

We note, however, that further viable models of chain inflation may still exist outside the validity of the bounce formalism. One may e.g.\ speculate, whether fast tunneling consistent with~\eqref{eq:rateconstraint} can also be realized through Hawking-Moss instantons~\cite{Hawking:1981fz}. We leave this and related questions for future work.

\section{Conclusion}

Chain inflation successfully describes the early expansion of the universe. It resolves the problem of non-percolating bubbles which plagues the ``old inflation'' proposal. In this work we investigated, whether chain inflation is consistent with the anisotropies observed in the CMB. For this purpose, we derived its power spectrum of primordial scalar fluctuations (see~\eqref{eq:powerfinal}). In contrast to slow roll inflation, the origin of fluctuations (studied in this paper) lies in the stochastic nature of tunneling which induces fluctuations in the number of vacuum transitions and, hence, in the inflaton field value.  We note a second mechanism (not studied in this paper) for perturbation generation due to bubble collisions of the final phase
transition, a subject for future work.
Our result proves that the observed amplitude of CMB fluctuations can indeed be generated in chain inflation,
due to the randomness of tunneling. A rather large tunneling rate per Hubble four-volume of $\Gamma/H^4\sim 10^{18}$ is required which translates to $\sim 10^4$ phase transitions per e-folding of inflation in the range of scales observable in the CMB.

This finding is model-independent and does not rely on any assumptions on the calculation of the tunneling rate.
Our result also helps to resolve a major confusion in the literature by confirming a previous calculation of Feldstein \& Tweedie~\cite{Feldstein:2006hm} while disputing a number of others. We, furthermore, derive a new expression for the scalar spectral index of chain inflation and show that the observed value $n_s\simeq 0.96$ can be realized if Hubble and tunneling rate are slowly varying functions of time. While a strong time-dependence of $\Gamma$ and/or $H$ is disfavored in the window of e-folds accessible to the CMB, it is less constrained outside this regime.

The strong instability of de Sitter vacua required by the CMB normalization turns out to be favorable from a quantum gravity perspective. It restricts chain inflation to a regime in which the Trans-Planckian Censorship Conjecture is automatically satisfied. This is remarkable in the light that slow roll inflation is facing pushback from similar Swampland conjectures. We thus consider it an extremely interesting question for future work whether concrete string theory realizations of chain inflation can be constructed. Particular promising to us appear generalizations of natural inflation~\cite{Freese:1990rb}, e.g. axion monodromy, modulated natural inflation and winding inflation. These have been proposed in the context of slow roll inflation but could also feature regimes of successful chain inflation.

As a first step towards the model realization we studied some generic properties of chain inflation with axions. In particular, we found a new approximation of the tunneling rate in periodic potentials by numerically solving the differential equation of the bounce. Our estimate replaces the thin-wall approximation in the phenomenologically most interesting regime of fast tunneling. By applying the CMB constraints as well as some basic arguments of quantum field theory we were able to derive an upper limit on the axion decay constant $f< 3\times 10^{10}\gev$ in axionic chain inflation. This constraint may pose a model building challenge for string theory, where axion decay constants typically come out larger. On the other hand, several mechanisms to lower the decay constant of string axions are known. We also emphasize that viable chain inflation models may exist outside the validity of the Coleman bounce -- tunneling via Hawking-Moss instantons being an example.

\section*{Acknowledgments}
K.F.\ is Jeff \& Gail Kodosky Endowed Chair in Physics at the University of Texas at Austin, and K.F.\ and M.W.\ are grateful for support via this Chair.  K.F.\ and M.W.\ acknowledge support by the Swedish Research Council (Contract No. 638-2013-8993).  K.F.\ acknowledges support from the U.S. Department of Energy, grant DE-SC007859.

\bibliography{chain}

\begin{thebibliography}{10}

\bibitem{Guth:1980zm}
A.~H. Guth,
\newblock Phys. Rev. D {\bf 23}, 347 (1981).

\bibitem{Guth:1982pn}
A.~H. Guth and E.~J. Weinberg,
\newblock Nucl. Phys. B {\bf 212}, 321 (1983).

\bibitem{Linde:1981mu}
A.~D. Linde,
\newblock Adv. Ser. Astrophys. Cosmol. {\bf 3}, 149 (1987).

\bibitem{Albrecht:1982wi}
A.~Albrecht and P.~J. Steinhardt,
\newblock Adv. Ser. Astrophys. Cosmol. {\bf 3}, 158 (1987).

\bibitem{Adams:1990ds}
F.~C. Adams and K.~Freese,
\newblock Phys. Rev. D {\bf 43}, 353 (1991), hep-ph/0504135.

\bibitem{Linde:1990gz}
A.~D. Linde,
\newblock Phys. Lett. B {\bf 249}, 18 (1990).

\bibitem{Freese:2004vs}
K.~Freese and D.~Spolyar,
\newblock JCAP {\bf 07}, 007 (2005), hep-ph/0412145.

\bibitem{Freese:2005kt}
K.~Freese, J.~T. Liu, and D.~Spolyar,
\newblock Phys. Rev. D {\bf 72}, 123521 (2005), hep-ph/0502177.

\bibitem{Freese:2006fk}
K.~Freese, J.~T. Liu, and D.~Spolyar,
\newblock (2006), hep-th/0612056.

\bibitem{Ashoorioon:2008pj}
A.~Ashoorioon, K.~Freese, and J.~T. Liu,
\newblock Phys. Rev. D {\bf 79}, 067302 (2009), 0810.0228.

\bibitem{Turner:1992tz}
M.~S. Turner, E.~J. Weinberg, and L.~M. Widrow,
\newblock Phys. Rev. D {\bf 46}, 2384 (1992).

\bibitem{Watson:2006px}
S.~Watson, M.~J. Perry, G.~L. Kane, and F.~C. Adams,
\newblock JCAP {\bf 11}, 017 (2007), hep-th/0610054.

\bibitem{Feldstein:2006hm}
B.~Feldstein and B.~Tweedie,
\newblock JCAP {\bf 04}, 020 (2007), hep-ph/0611286.

\bibitem{Huang:2007ek}
Q.-G. Huang,
\newblock JCAP {\bf 05}, 009 (2007), 0704.2835.

\bibitem{Chialva:2008zw}
D.~Chialva and U.~H. Danielsson,
\newblock JCAP {\bf 10}, 012 (2008), 0804.2846.

\bibitem{Chialva:2008xh}
D.~Chialva and U.~H. Danielsson,
\newblock JCAP {\bf 03}, 007 (2009), 0809.2707.

\bibitem{Cline:2011fi}
J.~M. Cline, G.~D. Moore, and Y.~Wang,
\newblock JCAP {\bf 08}, 032 (2011), 1106.2188.

\bibitem{Coleman:1977py}
S.~R. Coleman,
\newblock Phys. Rev. D {\bf 15}, 2929 (1977),
\newblock [Erratum: Phys.Rev.D 16, 1248 (1977)].

\bibitem{Bedroya:2020rac}
A.~Bedroya, M.~Montero, C.~Vafa, and I.~Valenzuela,
\newblock (2020), 2008.07555.

\bibitem{Easther:2009ft}
R.~Easther, J.~Giblin, John~T., L.~Hui, and E.~A. Lim,
\newblock Phys. Rev. D {\bf 80}, 123519 (2009), 0907.3234.

\bibitem{Giblin:2010bd}
J.~Giblin, John~T., L.~Hui, E.~A. Lim, and I.-S. Yang,
\newblock Phys. Rev. D {\bf 82}, 045019 (2010), 1005.3493.

\bibitem{Bunn:1996py}
E.~F. Bunn, A.~R. Liddle, and M.~J. White,
\newblock Phys. Rev. D {\bf 54}, 5917 (1996), astro-ph/9607038.

\bibitem{Akrami:2018odb}
Planck, Y.~Akrami {\em et~al.},
\newblock Astron. Astrophys. {\bf 641}, A10 (2020), 1807.06211.

\bibitem{Bedroya:2019snp}
A.~Bedroya and C.~Vafa,
\newblock JHEP {\bf 09}, 123 (2020), 1909.11063.

\bibitem{Duncan:1992ai}
M.~J. Duncan and L.~G. Jensen,
\newblock Phys. Lett. B {\bf 291}, 109 (1992).

\bibitem{Abbott:1984qf}
L.~Abbott,
\newblock Phys. Lett. B {\bf 150}, 427 (1985).

\bibitem{Silverstein:2008sg}
E.~Silverstein and A.~Westphal,
\newblock Phys. Rev. D {\bf 78}, 106003 (2008), 0803.3085.

\bibitem{McAllister:2008hb}
L.~McAllister, E.~Silverstein, and A.~Westphal,
\newblock Phys. Rev. D {\bf 82}, 046003 (2010), 0808.0706.

\bibitem{Flauger:2009ab}
R.~Flauger, L.~McAllister, E.~Pajer, A.~Westphal, and G.~Xu,
\newblock JCAP {\bf 06}, 009 (2010), 0907.2916.

\bibitem{Kim:2004rp}
J.~E. Kim, H.~P. Nilles, and M.~Peloso,
\newblock JCAP {\bf 01}, 005 (2005), hep-ph/0409138.

\bibitem{Kappl:2015esy}
R.~Kappl, H.~P. Nilles, and M.~W. Winkler,
\newblock Phys. Lett. B {\bf 753}, 653 (2016), 1511.05560.

\bibitem{Winkler:2019hkh}
M.~W. Winkler, M.~Gerbino, and M.~Benetti,
\newblock Phys. Rev. D {\bf 101}, 083525 (2020), 1911.11148.

\bibitem{Hebecker:2015rya}
A.~Hebecker, P.~Mangat, F.~Rompineve, and L.~T. Witkowski,
\newblock Phys. Lett. B {\bf 748}, 455 (2015), 1503.07912.

\bibitem{Hebecker:2018fln}
A.~Hebecker, D.~Junghans, and A.~Schachner,
\newblock JHEP {\bf 03}, 192 (2019), 1812.05626.

\bibitem{Baacke:2003uw}
J.~Baacke and G.~Lavrelashvili,
\newblock Phys. Rev. D {\bf 69}, 025009 (2004), hep-th/0307202.

\bibitem{Dunne:2005rt}
G.~V. Dunne and H.~Min,
\newblock Phys. Rev. D {\bf 72}, 125004 (2005), hep-th/0511156.

\bibitem{Svrcek:2006yi}
P.~Svrcek and E.~Witten,
\newblock JHEP {\bf 06}, 051 (2006), hep-th/0605206.

\bibitem{Ibanez:2015fcv}
L.~E. Ibanez, M.~Montero, A.~Uranga, and I.~Valenzuela,
\newblock JHEP {\bf 04}, 020 (2016), 1512.00025.

\bibitem{Hawking:1981fz}
S.~Hawking and I.~Moss,
\newblock Adv. Ser. Astrophys. Cosmol. {\bf 3}, 154 (1987).

\bibitem{Freese:1990rb}
K.~Freese, J.~A. Frieman, and A.~V. Olinto,
\newblock Phys. Rev. Lett. {\bf 65}, 3233 (1990).

\end{thebibliography}
\bibliographystyle{h-physrev.bst}

\end{document}